\documentclass{ws-ijgmmp}
\usepackage{float}
\usepackage{hyperref}
\usepackage{mathtools}
\begin{document}

\markboth{S.H. Shekh, Simran Arora, V. R. Chirde, P.K. Sahoo}
{Thermodynamical aspects of relativistic hydrodynamics in $f(R,G)$ gravity}

\catchline{}{}{}{}{}
%
%%%%%%%%%%%%%%%%%%%%%%%%%%%%%%%%%%%%%%%%%%%%%%%%%%%%%%%%%%%%%%%%%%%%

\title{\textbf{Thermodynamical aspects of relativistic hydrodynamics in $f(R,G)$ gravity}}

\author{S. H. Shekh}

\address{Department of Mathematics, S.P.M. Science and Gilani Arts and Commerce College,\\ Ghatanji, Yavatmal, Maharashtra-445301, India.\\
\email{da\_salim@rediff.com}}

\author{Simran Arora}

\address{Department of Mathematics,\\ Birla Institute of Technology and Science-Pilani,
Hyderabad Campus,\\
Hyderabad-500078, India\\
\email{dawrasimran27@gmail.com}}

\author{V. R. Chirde}

\address{Department of Mathematics, G.S.G. Mahavidyalaya, Umarkhed-445206, India.\\
\email{vrchirde333@rediffmail.com}}

\author{P.K. Sahoo}

\address{Department of Mathematics,\\ Birla Institute of Technology and Science-Pilani,
Hyderabad Campus,\\
Hyderabad-500078, India,\\
\email{pksahoo@hyderabad.bits-pilani.ac.in} }

\maketitle

\begin{history}
\received{(27 Nov 2019)}
\accepted{(26 Jan 2020)}
\end{history}

\begin{abstract}
Present investigation devoted to the dynamical study of Relativistic Hydrodynamics with some thermodynamical characteristics in $f(R,G)$ gravity towards spatially homogeneous isotropic cosmological model filled with isotropic fluid. We govern the features of the derived cosmological model by considering the power-law inflation for the average scale factor. The temperature and entropy density of the proposed model are positive definite. We also discuss  the energy conditions to our solutions. The strong energy condition violated, which indicates the accelerated expansion of the proposed model.
\end{abstract}



\section{Introduction}\label{I}

The relativistic hydrodynamics is about the physical properties of fluids in which either the bulk viscosity of the flow is comparable with the speed of light or the intensity of the gravitational field which is either the background or generated by matter itself or when the space-time curvature is large. In fact, the application of work ranges from astrophysical phenomenon to a relativistic treatment. A relativistic description is significant topic of astrophysics in the following ways: (a) jets emanating at relativistic speed from the core of active galactic nuclei, (b) in framework which involves gravitational collapse of compact stars and flows around black holes. (c) Montero et al. \cite{1} report on the usage of relativistic hydrodynamics, uniting with dynamical space-times, in spherical polar coordinates without symmetry assumptions. They employ a high-resolution shock-capturing scheme, which requires that the equations be cast in flux-conservative form while Font \cite{2} review formulations of the equations of general relativistic hydrodynamics and magneto hydrodynamics, along with methods for their numerical solution.

Recent observational studies, which includes the supernovae cosmology project \cite{3,4}, have provided the major indication for the cosmic acceleration of the universe along with some observations like those of the distant supernovae, large scale structure (LSS) \cite{5,6} fluctuations of the cosmic microwave background radiation, the Wilkinson microwave anisotropy probe (WMAP) \cite{7}, the Sloan Digital Sky Survey (SDSS) \cite{8}, and the Chandra X-ray observatory \cite{9} suggest that our universe is undergoing an accelerated expansion. The evidence that has been increased to explain this observed phenomenon can be classified into two categories. First, within the framework of Einstein's General Theory of Relativity (GTR), an exotic component filled with negative pressure called mysterious energy or Dark Energy (DE) which has been conventionally characterized by the equation of state (EoS) parameter $\omega=\frac{p}{\rho}$ that is not necessarily constant. Oli \cite{10} presented some cosmological parameters during the evolution of the universe for the class of solutions of the field equations describing two-fluid universe for the Bianchi type-I model considering one radiating and other matter content fluid in interacting and non-interacting scenarios. 

Next alternative is to modify the action of GTR theory called as Modified Theory of Gravity (MTG). Recently, research in cosmology has seen a growing interest in theories of gravity beyond GTR. Many models of MTG have been introduced in order to tackle the shortcomings of GTR such as well-known $f(R)$ gravity which replaces the Ricci scalar in the action by an arbitrary function of Ricci scalar. Capozziello et al. \cite{11} obtained dust matter and dark energy phase of the universe using power law cosmology in $f(R)$ gravity. Later Azadi et al. \cite{12} studied vacuum solution in cylindrically symmetric space-time in $f(R)$ gravity. Chirde and Shekh \cite{13} are the authors who have investigated an interaction between the barotropic fluid and dark energy with a zero-mass scalar field in $f(R)$ gravity for the spatially homogeneous and isotropic flat FRW universe. They have used  the volumetric power law and exponential law of expansion for the plane-symmetric cosmological model along with the quadratic equation of state in the metric version of $f(R)$ gravity by allowing negative constant deceleration parameter. Several authors have inspected the aspects of some cosmological models in this gravity \cite{14,15,16,17}. Amongst the various adaptations of Einstein's theory, another one way to look at the theory beyond GTR is the Teleparallel Gravity (TG) in which the Weitzenbock connection is used in place of the Levi-Civita connection and therefore, it has no curvature but has torsion which takes the responsibility of the acceleration of the universe. Some relevant works in this gravity are presented in \cite{18,19,20,21,22}. In \cite{18} the authors described the graphical representation of k-essence using EoS parameter. Wang \cite{19} obtained some spherically symmetric solutions. Bohmer et al. \cite{20} investigated the existence of relativistic stars in gravity whereas some cosmological models with different sources have been discussed by Chirde and Shekh \cite{21}.  Recently, the discussion of stability of the accelerating universe using a linear EoS in $f(T)$ gravity with hybrid expansion law is given by Bhoyar et al. \cite{22}.  

Recent developments in $f(R)$ gravity uses Lovelock invariants, such as the Gauss-Bonnet scalar $G$ and some works that can successfully describe the dark energy era and also the inflationary era theoretically. $f(R)$ and $f(G)$ gravity have generalizations offered by higher-order gravities which use combinations of higher-order curvature invariants constructed from the Ricci and Riemann tensors. Also the theory which combines Ricci scalar and Gauss-Bonnet scalar called $f(R,G)$ gravity theory. The Gauss-Bonnet term is added first time to Einstein action as gravitational dark energy in Ref \cite{Nojiri}. Later, the dark energy and ghost free Gauss-Bonnet modified gravity theories are discussed in detail \cite{Cognola,SNojiri}. Alvaro de la Cruz-Dombriz and Diego SáezGómez \cite{23} focused on the analysis of $f(R,G)$ gravity and a deep analysis has been performed on the stability of some important cosmological solutions which not only convince to constrain the form of the gravitational action, but also further help in better understanding of the perturbations behaviour in the higher-order theories of gravity. This will lead to a more precise analysis of the full spectrum of cosmological perturbations. The existence of $f(R,G)$ gravity by Noether symmetries approach is discussed in  \cite{24}.  The authors in \cite{24} derived the exact solutions by the reduction of cosmological dynamical system and the presence of conserved quantities. Costa et al. \cite{25} used a dynamical system approach to discuss the cosmological viability of $f(R,G)$ gravity theories. Benetti et al. \cite{26} have discussed the observational existence of power law solutions for a class of $f(R,G)$ models derived using Noether’s symmetries. They have concluded that the used geometrical description with power law solutions can describe the current available observational data without contributing dark energy in $f(R,G)$ gravity. By taking dark components of the cosmological Hubble flow, Capozziello et al. \cite{27} illustrated that any analytic theory of $f(R,G)$ gravity for n-dimensional FRW metric can be associated to a perfect-fluid stress-energy tensor.

Recently, Shekh and Chirde \cite{28} have investigated the plane symmetric cosmological model in different theories of gravitation namely GTR, $f(R)$ and $f(T)$ gravity with hydrodynamic source. It is observed that the fluid in GTR is fully occupied with quintessence dark energy fluid whereas the model shows both matter and dark energy dominated era in $f(R)$ gravity and remains present in matter dominated era while in $f(T)$ gravity. The model initially shows standard Cold Dark Matter (CDM) model and at the expansion it is fully occupied with quintessence dark energy fluid. The present article is organized in sections with some basis of $f(R,G)$ gravity with relativistic hydrodynamic sources in Sec. II. In Sec. III, we classify the models based on deceleration parameter and Hubble parameter. Thermodynamical behavior and entropy of the model is derived in Sec. IV. In Sec. V we derived the detail solutions of the FRW metric along with thermodynamical temperature and entropy density. Finally, we summarize our work in Sec. VI. We considered $8\pi \mathcal{G}=c=1$ throughout the work.

\section{Basics of $f(R,G)$ gravity with Hydrodynamic source}\label{II}

The most general action for $f(R,G)$ gravity is given as \cite{23}
\begin{equation}\label{e1}
S=\frac{1}{2k}\int d^{4}x\sqrt{-g}(R+f(G))+ S_{M}(g^{ij},\varphi),
\end{equation}
where $S_{M}(g^{ij},\varphi)$ is the matter action, $R$ is Ricci scalar and $G$ is Gauss-Bonnet invariant defined by 
\begin{equation}\label{e2}
G=R^{2}-4R_{\alpha\beta}R^{\alpha \beta}+R_{\alpha \beta \sigma \nu}R^{\alpha \beta \sigma \nu},
\end{equation}
where, the notations $R_{\alpha\beta}$ and $R_{\alpha \beta \sigma \nu}$ are occupied for the Ricci and Riemann tensors respectively.

Variation of the standard action (\ref{e1}) with respect to the metric gives us the following gravitational field equation:

\begin{equation}\label{e3}
R_{\mu \nu}-\frac{1}{2}g_{\mu \nu}R=k T_{\mu \nu}^{mat}+\Sigma_{\mu \nu},
\end{equation}
where,
\begin{multline}\label{e4}
\Sigma_{\mu \nu}=\nabla_{\mu}\nabla_{\nu}f_{R}-g_{\mu \nu}\square f_{R}+2R\nabla_{\mu}\nabla_{\nu}f_{G}-2g_{\mu \nu}\square f f_{G}-4R^{\lambda}_{\mu}\nabla_{\lambda}\nabla_{\nu}f_{G}-4R^{\lambda}_{\nu}\nabla_{\lambda}\nabla_{\mu}f_{G}+4\square_{\mu \nu}f_{G}+
\\4g_{\mu \nu}R^{\alpha \beta}\nabla_{\alpha}\nabla_{\beta}f_{G}+4R_{\mu \alpha \beta \nu}\nabla^{\alpha}\nabla^{\beta}f_{G}-\frac{1}{2}g_{\mu \nu}(f_{R}R+f_{G}G-f(R,G))+(1-f_{R})(R_{\mu \nu}-\frac{1}{2}g_{\mu \nu}R).
\end{multline}
Here, $\nabla_{\mu}$ represents the covariant derivative.
\begin{equation}\label{e5}
f_{R}\equiv\frac{\partial f(R,G)}{\partial R}  \hspace{2mm} and \hspace{2mm} f_{G}\equiv\frac{\partial f(R,G)}{\partial G},
\end{equation}
gives the partial derivatives of $f(R,G)$ with respect to $R$ and $G$ respectively.

Makarenko et al. \cite{29} established the cosmological reconstruction in $f(R,G)$ gravity and obtained the phantom type cosmological model which do not lead to a future singularity. Atazadeh and Darabi \cite{30} studied the viability of $f(R,G)$ gravity by imposing energy conditions using two forms of $f(R,G)$, accounting for the stability of cosmological solutions and also constructed the inequalities obtained by energy conditions. After that individually applying the weak energy condition using the recent estimated values of the  deceleration, Hubble, jerk and snap parameters to probe the viability. Laurentis and Lopez-Revelles \cite{31} discuss the detail investigation of the weak field limit of $f(R,G)$ gravity taking into consideration an analytic functions of the Ricci scalar $R$ and the Gauss-Bonnet invariant $G$, specifically by developing in metric formalism, the Newtonian, Post- Newtonian and Parametrized Post-Newtonian limits starting from general $f(R,G)$ Lagrangian and observed the Newtonian limit of $f(R,G)$ gravity. Shamir and Zia \cite{32} highlighted the materialization of anisotropic compact stars namely Her X1, SAX J 1808-3658, and 4U 1820-30 in the context of $f(R,G)$ theory of gravity and have shown that all three stars behave as usual as for positive values of the $f(G)$ model parameter $n$. In this work, we obtain the solution of field equations and the behavior of the universe using some kinematical and physical quantities for the $f(R,G)$ gravity model i.e.
\begin{equation}\label{e6}
f(R,G)=f_{0}R^{m}G^{1-m},
\end{equation}
where $f_{0}>0$ be any constant.

For the values of constant $m$,  two types of gravity models are recovered i) $f(R)$ gravity model corresponding to  $f_{0}=1$ and $m=1$ while ii) $f(G)$ gravity model corresponding to $f_{0}=1$ and $m=0$.

The General-Relativistic Hydrodynamics (GRH) equations consist of the local conservation laws of the matter current density, $J^{\mu}$ (the continuity equation) and of the stress-energy tensor, $T^{\mu \nu}$ (the Bianchi identities):
\begin{equation}\label{e7}
\nabla_{\mu}J^{\mu}=0 ,\nabla_{\mu}T^{\mu \nu}=0,
\end{equation}
where usual $\nabla_{\mu}$ stands for the covariant derivative associated with the four-dimensional space-time metric $g_{\mu \nu}$. The density current is given by $J^{\mu}=\rho u^{\mu},u^{\mu}$ represent the fluid 4-velocity and $\rho$ the proper rest-mass density.

The stress-energy momentum tensor $T_{\mu \nu}$ for a non-perfect (un-magnetized) fluid is defined as 
\begin{equation}\label{e8}
T_{\mu \nu}=\rho(1+\varepsilon)u_{\mu}u_{\nu}+(p-\zeta\theta)h_{\mu \nu}-2\eta\sigma_{\mu \nu}+q_{\mu}u_{\nu}+q_{\nu}u_{\mu},
\end{equation}
where $\varepsilon$ is the specific energy density of the fluid in its rest frame, $p$ is the pressure, and $h_{\mu\nu}$ is the spatial projection tensor $h_{\mu\nu}=u_{\mu}u_{\nu}+g_{\mu \nu}$. In addition, $\zeta$  and $\eta$ are the shear scalar and bulk viscosities. The scalar expansion $\theta$, describe the convergence or divergence of the fluid world lines and finally, $q_{\mu \nu}$ is the energy flux vector. 

In the following, we will be neglecting non-adiabatic effects, like viscosity and heat transfer, considering the stress-energy tensor to be a perfect fluid
\begin{eqnarray}\label{e9}
T_{\mu \nu}=\rho hu_{\mu}u_{\nu}-pg_{\mu \nu},\\
T_{11}=T_{22}=T_{33}=-p, T_{44}=\rho h,
\end{eqnarray}
where $h$ is the relativistic specific enthalpy and is defined by
\begin{equation}\label{e10}
h=1+\varepsilon+\frac{p}{\rho},
\end{equation}

Now, in order to close the system, the equation of motion and the continuity equation must be supplemented with an EoS relating some fundamental thermodynamic quantities. In general, the EoS takes the form
\begin{equation}\label{e11}
p=p(\rho,\varepsilon),
\end{equation}
The available EoS has become sophisticated enough to take into account the physical and chemical processes such as Quantization, Molecular Interactions, Nuclear Physics, Relativistic Effects,  etc. However, due to their simplicity, the most widely occupied EoS in numerical simulations in astrophysics is the ideal fluid EoS,
\begin{equation}\label{e12}
p=(\Xi-1)\rho\varepsilon,
\end{equation}
where $\Xi$ is known as adiabatic index.

The Polytropic EoS (e.g., to build equilibrium stellar models),
\begin{equation}\label{e13}
p=K \rho^{\Xi}\thickapprox K \rho^{1+\frac{1}{N}},
\end{equation}
where $N$ be the Polytropic index \& $K$ is the Polytropic constant and the Microphysical EoS that describe the interior of compact stars at nuclear matter densities have also been developed.

Using equations (\ref{e8}) and (\ref{e9}), the equation (\ref{e6}) becomes
\begin{equation}\label{e14}
h=1+\Xi \varepsilon,
\end{equation}
and
\begin{equation}\label{e15}
h=1+\varepsilon+K \rho^{\frac{1}{N}},
\end{equation}

\section{Explication of Isotropic Homogeneous Space-Time}\label{III}

One can classify models of the universe on the basis of the time dependence of the deceleration  and Hubble's parameter. Both the parameters can change their sign during the evolution of the universe. Therefore the evolving universe can transit from one type to another. It is one of the basic tasks of cosmology to follow this evolution and clarify its causes. When the Hubble's parameter is constant, the deceleration parameter is also constant and equal to -1, as in the de-Sitter and steady-state universe. All models can be characterized as follows whether they accelerate or decelerate, and expand or contract :\\
(a) $q>0, H>0:$ the model is decelerating and expanding\\
(b) $q<0, H>0:$ the model is accelerating and expanding\\
(c) $q>0, H<0:$ the model is decelerating and contracting\\
(d) $q<0, H<0:$ the model is accelerating and contracting\\
(e) $q=0, H>0:$ the model is expanding with no deceleration\\
(f) $q=0, H<0:$ the model is contracting with no deceleration\\
(g) $q=0, H=0:$ the model is static.\\
From the above only (a), (b), and (e) are possible. But the evidence in favor of the fact that the expansion is presently accelerating continuously grows in number and therefore the current dynamics belongs to type (b).\\
According to Berman \cite{33, 34}, the deceleration parameter and scale factor are associated by the relation
\begin{equation}\label{e16}
q=-\frac{a\ddot{a}}{\dot{a}^{2}}.
\end{equation}
where $a$ be the scale factor and $q$ be the deceleration parameter.

The sign of $q$ indicates whether the model will inflate or not. If $q$ is negative then the model indicates inflation. Also, recent observations of type Ia supernovae, reveal that the present universe is accelerating and the value of deceleration parameter lies somewhere in the range $-1\leq q \leq 0$.
The deceleration parameter can be constant if we relate the Hubble's parameter $H$ to the scale factor $a$,
\begin{equation}\label{e17}
H=ba^{-m}=bV^{\frac{-m}{3}},
\end{equation}
where $b$ and $m$ are constants.

Using equation (\ref{e16}), we can re-write the above equation as
\begin{equation}\label{e18}
\dot{a}=ba^{-m+1},
\end{equation}
\begin{equation}\label{e19}
\ddot{a}=-b^{2}(m-1)a^{-2m+1}.
\end{equation}
Using equations (\ref{e16}), (\ref{e18}) and (\ref{e19}), we get
\begin{equation}\label{e20}
q=-1+m.
\end{equation}
This equation gives a constant value for deceleration parameter and it can take both positive as well as negative values. Positive value of deceleration parameter results into the standard deceleration model while the negative value results into inflation or an accelerating model.
On solving equation (\ref{e16}) we get
\begin{equation}\label{e21}
a=(\alpha t+\beta)^{\gamma}, \hspace{2mm} \text{where} \hspace{2mm} \gamma=\frac{1}{1+q} \hspace{2mm} \text{and} \hspace{2mm} q\neq-1.
\end{equation}
Provided $\alpha\neq0$ and $\beta$ are constants of integration.\\
From the equation (\ref{e21}), it is observed that the average scale factor of the model is the function of cosmic time, which increase with time at $q>-1$, decreases with time at $q<-1$, and does not exist at $q=-1$. Also, it is observed that these parameters start with a constant value for $q>-1$, except the point $t_{s}=\frac{-\beta}{\alpha}$, for this point $t_{s}$ it starts with zero hence the model has singularity (point type) \cite{35} at the point $t_{s}$.

\section{Thermodynamical behavior and entropy of the model}\label{IV}

Thermodynamical analysis has become a powerful tool to inspect a gravitational theory. As pivotal events, black hole thermodynamics and recent AdS (Anti-de Sitter Space) /CFT (Conformal Field Theory) correspondence show the explicit significance and strongly suggest the deep connection between gravity and thermodynamics.

From the thermodynamics, we apply the combination of the first and second law of thermodynamics to the system with volume $V$ \cite{13},
\begin{equation}\label{e22}
\tau ds=d(\rho V)+\rho dV,
\end{equation}
where $\tau$ and $s$ represent the temperature and entropy respectively.

Above equation may be written as 
\begin{equation}\label{e23}
\tau ds=d[(p+\rho)V]-Vdp.
\end{equation}
The integrability condition is necessary to define a perfect fluid as a thermodynamic system; it is given by
\begin{equation}\label{e24}
dp=\bigg(\frac{p+\rho}{\tau}\bigg) d\tau.
\end{equation}
Using equations (\ref{e23}) and (\ref{e24}) we have the differential equation
\begin{equation}\label{e25}
ds=\frac{1}{\tau} d[(p+\rho)V]-(p+\rho)V \frac{d\tau}{\tau^{2}}.
\end{equation}

Rewriting above equation
\begin{equation}\label{e26}
ds=d\bigg[\frac{(p+\rho)V}{\tau}\bigg].
\end{equation}

Therefore the entropy is defined as
\begin{equation}\label{e27}
s=\bigg[\frac{(p+\rho)V}{\tau}\bigg].
\end{equation}

Let the entropy density be $s'$, so that
\begin{equation}\label{e28}
s'=\frac{s}{V}=\bigg(\frac{p+\rho}{\tau}\bigg) =\frac{(1+\omega)\rho}{\tau}.
\end{equation}

If we define the entropy density in terms of temperature, then the first law of thermodynamics may be
written as
\begin{equation}\label{e29}
d(\rho V)+\omega\rho dV=(1+\omega)\tau d\bigg(\frac{\rho V}{\tau}\bigg),
\end{equation}

which on integration yields
\begin{equation}\label{e30}
\tau=\rho^{\frac{\omega}{1+\omega}}
\end{equation}

From equation (\ref{e28}), we obtain
\begin{equation}\label{e31}
s'=(1+\omega)\rho^{\frac{1}{1+\omega}}
\end{equation}

Equation (\ref{e27}) represents the thermodynamics of the universe (entropy) which does not depend on any individual fluids, it depends on the total matter density and isotropic pressure of the fluid. Chirde and Shekh \cite{13} are the author who have investigated the behavior of accelerating spatially homogeneous and isotropic Friedman-Robertson-Walker cosmological model with a different option of barotropic viscous fluid in the framework of some well-known $f(T)$ gravity model by defining some basic thermodynamic aspects such as, thermodynamic temperatures and entropy densities of the model with the help of a power-law solution. The remark on the actions of thermodynamic parameters  are directly related to the energy density of the universe. Hence our outcomes in equations (\ref{e30}) and (\ref{e31}) shows the same features with the work prepared by the above authors.

\section{Metric, field equation and their solutions }\label{V}
\subsection{Isotropic model}
We consider the spatially homogeneous and isotropic Friedman-Robertson-Walker (FRW) line element in the form
\begin{equation}\label{e32}
ds^{2}=dt^{2}-a^{2}(t)\biggl[\frac{dr^{2}}{1-kr^{2}}+r^{2}d\theta^{2}+r^{2}\sin^{2}\theta d\phi^{2}\biggr],
\end{equation}
where $a(t)$ be the scale factor of the universe.\\
The angle $\theta$ and $\phi$ are the usual azimuthal and polar angles of spherical coordinates, with $0\leq \theta \leq \phi$ and $0\leq \phi \leq \phi$. The coordinates $(t,r,\theta,\phi)$ are called comoving coordinates. This tells that the coordinate system follows the expansion of space so that the space coordinates of objects which do not move with respect to the background remain the same. The homogeneity of the universe fixes a special frame of reference, the cosmic rest frame given by the above coordinate system. Also, $k$ is a constant which represent the curvature of the space-time. If $k =1$, then this corresponds to a closed universe, the flat universe is obtained for $k=0$ and $k = -1$ corresponds to an open universe. In this work, we deliberate on the flat universe taken after $k=0$ with infinite radius.

The equation of motion (\ref{e4}) for the spatially homogeneous and isotropic FRW line element (\ref{e32}) with the fluid of stress-energy tensor can be written as
\begin{equation}\label{e33}
2f_{R}\frac{\dot{a^{2}}}{a^{2}} + 3\dot{f_{R}}\frac{\dot{a}}{a}+12\dot{f_{G}}\frac{\dot{a^{3}}}{a^{3}} -\frac{1}{2}(Rf_{R}+Gf_{G}-f)=\rho ,
\end{equation}

\begin{equation}\label{e34}
\ddot{f_{R}}+2\dot{f_{R}}\frac{\dot{a}}{a}+4\frac{\dot{a}}{a}\bigg(\frac{\dot{a}}{a}\ddot{f_{G}}+\frac{2 \ddot{a}}{a}\dot{f_{G}}\bigg)+f_{R}\bigg(\frac{2\ddot{a}}{a}+\frac{\dot{a^{2}}}{a^{2}}\bigg)-\frac{1}{2}(Rf_{R}+Gf_{G}-f)=-p.
\end{equation}
The overhead dot represents the differentiation with respect to time $t$.
Now, we consider some of the kinematical parameters for the FRW cosmological model that are important in cosmological observations.\\
The spatial volume,
\begin{equation}\label{e35}
V=a^{3},
\end{equation}
The generalized mean Hubble parameter,
\begin{equation}\label{e36}
H=\frac{\dot{a}}{a}.
\end{equation}
The mean anisotropy parameter,
\begin{equation}\label{e37}
A_{m}=\frac{1}{3}\sum^{3}_{i=1}\bigg(\frac{H_{i}-H}{H}\bigg)^{2}.
\end{equation}
The expansion scalar and shear scalar,
\begin{equation}\label{e38}
\theta=u_{;\mu}^{\mu}=\frac{\dot{A}}{A}+\frac{\dot{B}}{B}+\frac{\dot{C}}{C}=3H,
\end{equation}
\begin{equation}\label{e39}
\sigma^{2}=\frac{3}{2}H^{2}A_{m}.
\end{equation}

\subsection{Kinematical parameters of the model}

\begin{figure}[H]
   \centerline{\includegraphics[scale=0.4]{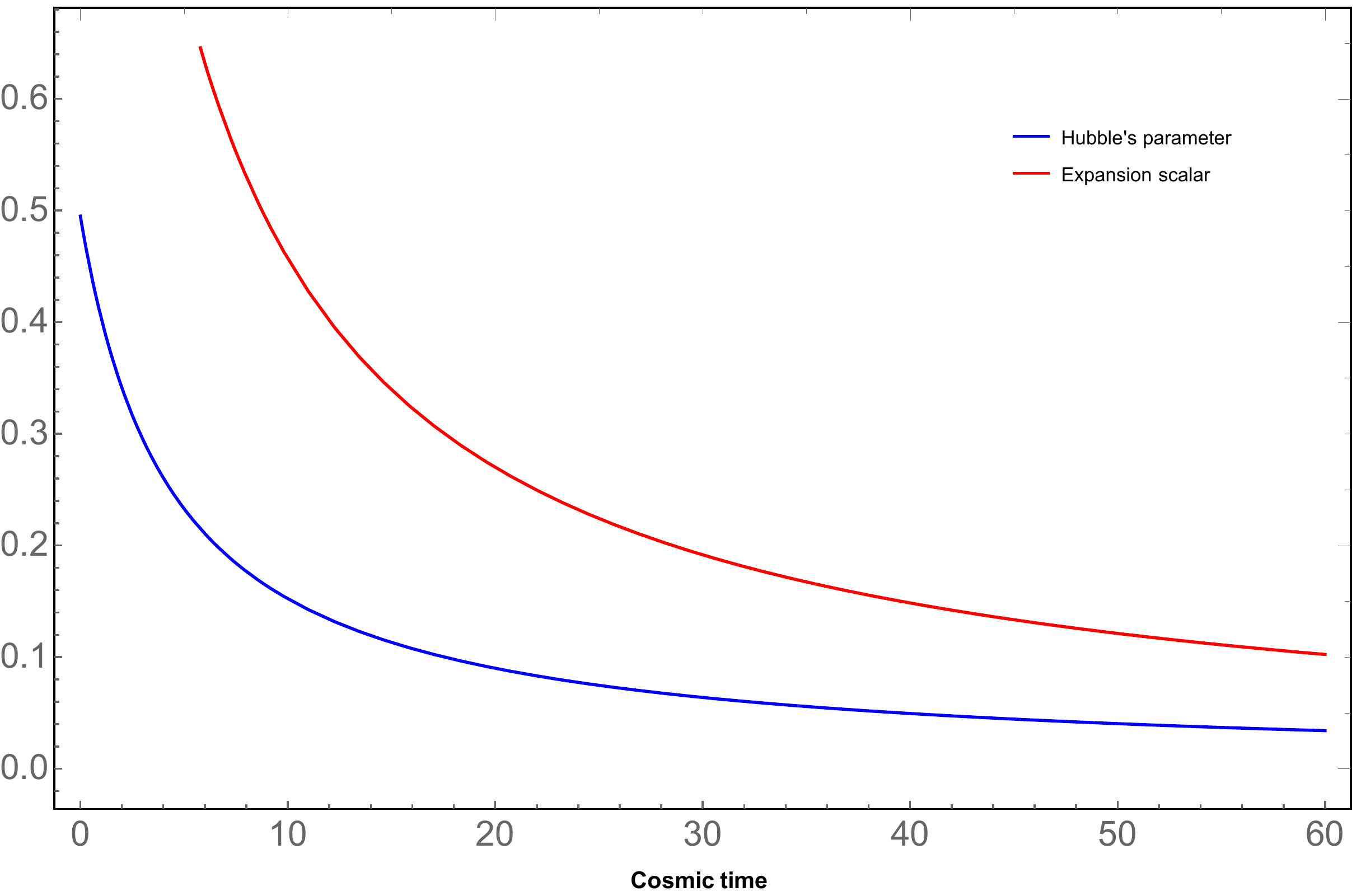}}
  \caption{Behavior of Hubble's parameter and Expansion scalar versus cosmic time $t$ with  $\alpha=0.9$, $\beta=4$, $\gamma=2.2$.}\label{Figure-i}
\end{figure}

\begin{figure}[H]
  \centerline{\includegraphics[scale=0.5]{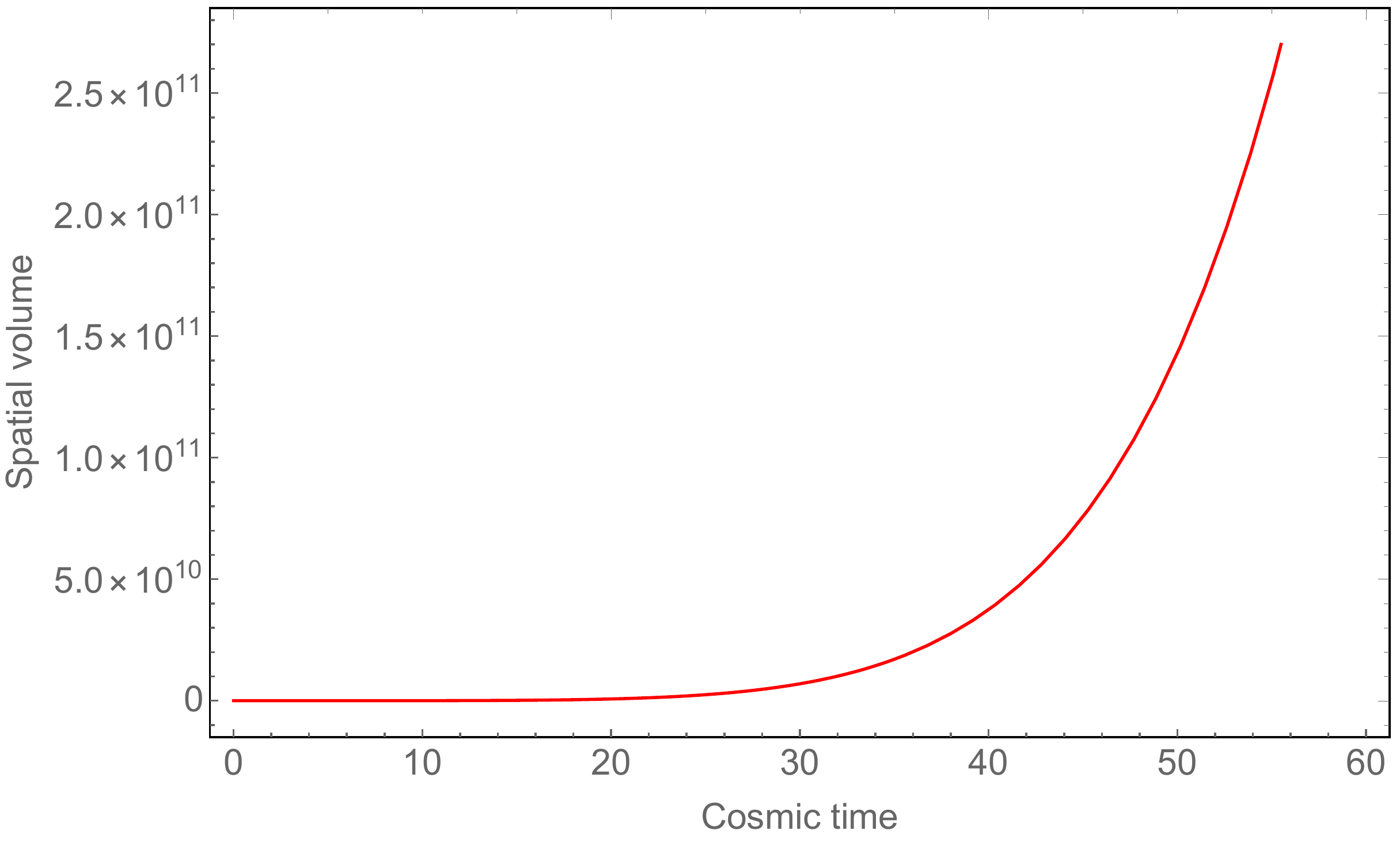}}
  \caption{Behavior of Spatial volume of the model versus cosmic time $t$ with $\alpha=0.9$, $\beta=4$, $\gamma=2.2$.}\label{Figure-ia}
\end{figure}

Using the value of scale factor given in equation (\ref{e21}), the kinematical parameters given in equations (\ref{e35}) to (\ref{e39}) are defined as:\\
The spatial volume,
\begin{equation}\label{e40}
V=(\alpha t+\beta)^{3\gamma}.
\end{equation}
In our analysis, we observed that the spatial volume $V$ of the universe starts with constant value as
$t\rightarrow 0$ and attains big-bang as $t\rightarrow\frac{-\beta}{\alpha}$ and also with the increase of cosmic time $t$ it always expands and increase. When $t\rightarrow \infty$ then spatial volume $V\rightarrow \infty$. Thus inflation is possible in flat FRW universe. This shows that the universe evolve with zero volume as $t\rightarrow 0$ and expands with cosmic time $t$. The behavior of spatial volume verses cosmic time $t$ is shown in Figure \ref{Figure-ia}.\\
The scalar expansion, 
\begin{equation}\label{e41}
\theta=\frac{3\alpha\gamma}{(\alpha t+\beta)}.
\end{equation}
The generalized Hubble parameter,
\begin{equation}\label{e42}
H=\frac{\alpha\gamma}{(\alpha t+\beta)}
\end{equation}
From the equation (\ref{e41}) and (\ref{e42}) it is observed that the expansion scalar and the generalized Hubble parameter is constant throughout the evolution of the universe as $t\rightarrow \infty$. This shows that the universe is expanding with the increase of cosmic time but the rate of expansion decrease to a constant value which shows that the universe starts evolving with zero volume as $t\rightarrow \infty$ with an infinite rate of expansion. The behavior of Hubble’s parameter and expansion scalar of the universe versus cosmic time $t$ is shown in Figure \ref{Figure-i}. The endeavors of the same are resembled with the work of \cite{35}.


\subsection{Physical parameters of the model}

We discussed the physical parameters of the model which described the physical interpretation of the model as follows:\\
Isotropic pressure of the model,
\begin{multline}\label{e43}
\displaystyle p=  \frac{m(m-1)[6\alpha^{2}\gamma(2\gamma-1)]^{m}}{2(\alpha t+\beta)^{2m}}-\frac{3m\alpha^{2}\alpha_{1}(3m^{2}+3\gamma^{2}+2m\gamma-7m-4\gamma+4)}{(\alpha t+\beta)^{5-3m}}\\
+\frac{4\alpha^{2}\gamma^{2}\alpha_{2}(2m^{2}-4m^{2}\gamma-3m\gamma^{2}-6m+7m\gamma+6\gamma^{2}-6\gamma)}{(\alpha t+\beta)^{4-2m}}.
\end{multline}

Energy density of the model,
\begin{multline}\label{e44}
\displaystyle \rho=\frac{m(1-m)[6\alpha^{2}\gamma(2\gamma-1)]^{m}}{2(\alpha t+\beta)^{2m}}-\frac{4m(\gamma-1)\alpha^{3}\alpha_{2}\gamma^{2}(2\alpha\gamma^{2}+9m-9)}{(2\gamma-1)(\alpha t+\beta)^{5-3m}}\\
-\frac{12\alpha^{3}\gamma^{3}\alpha_{2}[2m+m(\gamma-1)-2(\gamma-1)]}{(\alpha t+\beta)^{4-2m}}.
\end{multline}
where $\alpha_{1}=\bigg[\frac{2\gamma-1}{4(\gamma-1)\alpha^{2}\gamma^{2}}\bigg]^{m-1}$ and $\alpha_{2}=\bigg[\frac{2\gamma-1}{4(\gamma-1)\alpha^{2}\gamma^{2}}\bigg]^{m}$.

\begin{figure}[H]
  \centerline{\includegraphics[scale=0.5]{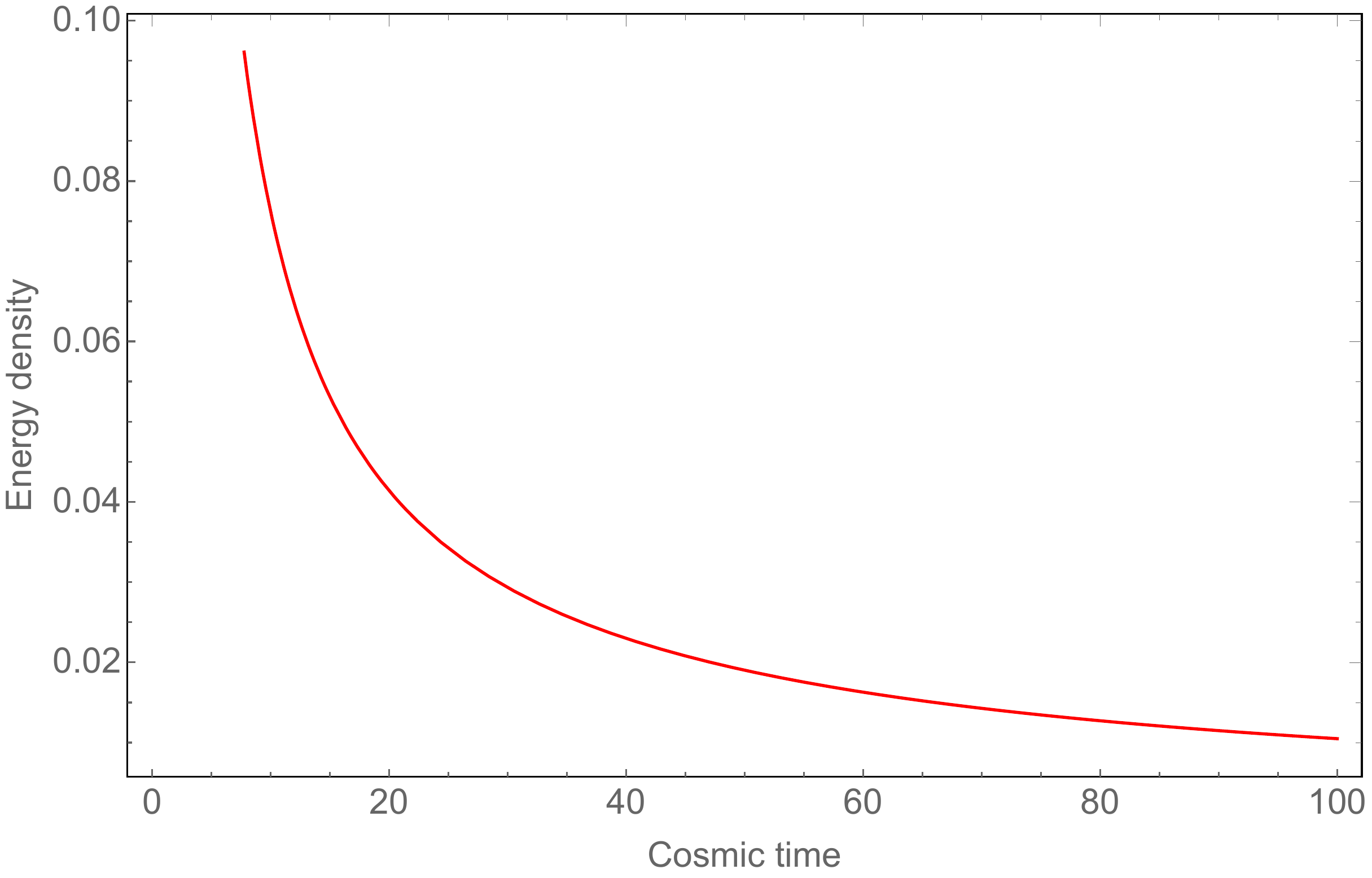}}
  \caption{Behavior of energy density of the model versus cosmic time $t$ with $\alpha=0.9$, $\beta=4$, $\gamma=2.2$ $m=0.45$, $\alpha_{1}=2.562$, $\alpha_{2}=0.463$.}\label{Figure-ii}
 \end{figure}
 \begin{figure}[H]
  \centerline{\includegraphics[scale=0.5]{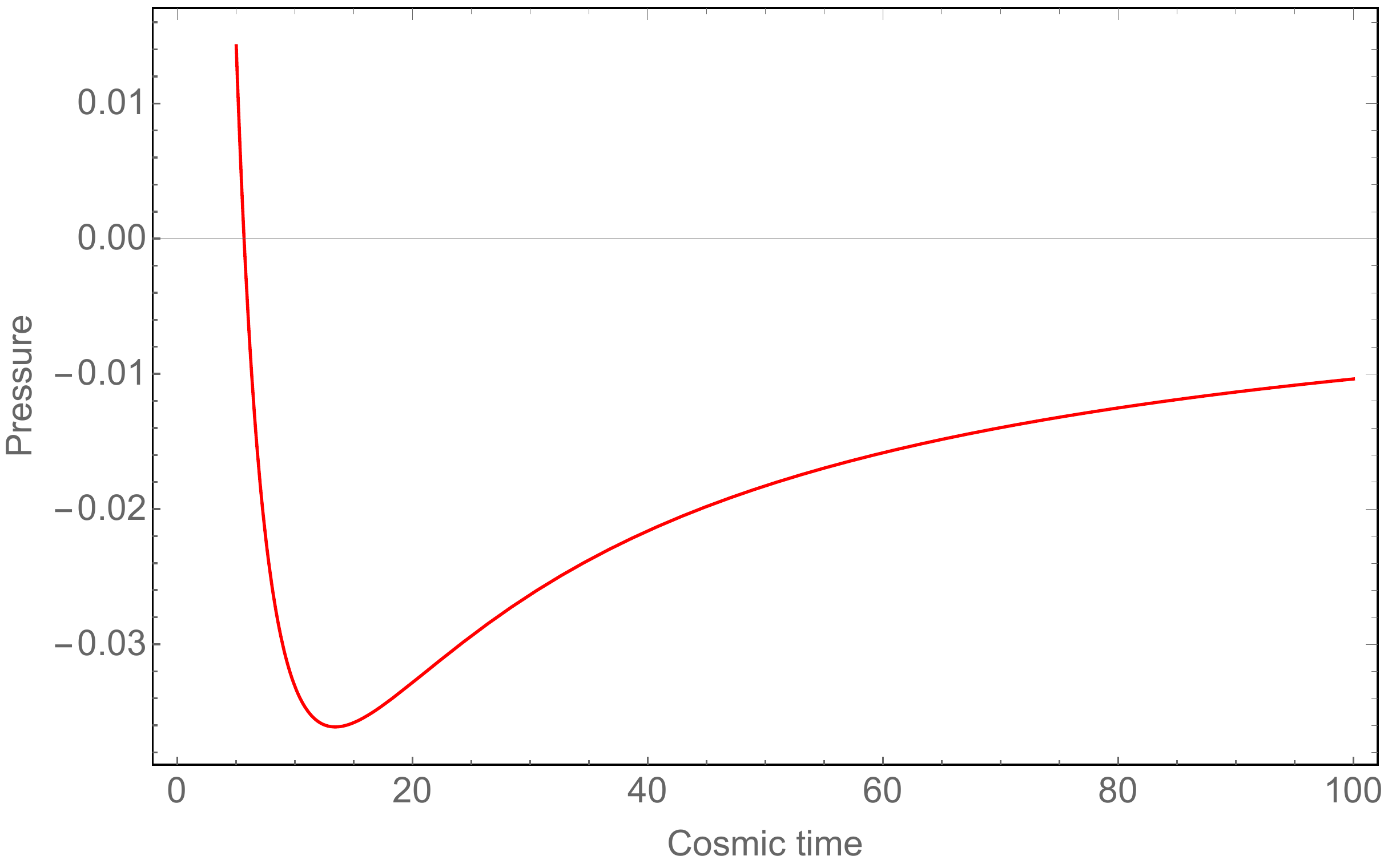}}
  \caption{Behavior of pressure of the model versus cosmic time $t$ with $\alpha=0.9$, $\beta=4$, $\gamma=2.2$ $m=0.45$, $\alpha_{1}=2.562$, $\alpha_{2}=0.463$.}\label{Figure-iia}
\end{figure}

Figures \ref{Figure-ii} and \ref{Figure-iia}, depict that the graphical variation of the energy density and isotropic pressure of the model versus cosmic time $t$ with the appropriate choice of constants. It is observed that the energy density distribution is positive decreasing function of time $t$. At the initial stage from where the model starts to expand i.e. when $t\rightarrow 0$ the energy density of the model is constant whereas with the expansion at $t>0$ it is infinite and at the infinite time ($t\rightarrow \infty$) it approaches to zero i.e. $\rho\rightarrow 0$, thus at infinite expansion the model is asymptotically empty whereas initially when universe start to expand for $0\leqslant t \leqslant 2.7 $ an isotropic pressure of the model is positive while with expansion it becomes negative hence the behavior  of the Universe is accelerated.

Equation of state parameter of the model,
\begin{equation}\label{e45}
\displaystyle \omega=\left[\dfrac{\splitdfrac{{\frac{m(m-1)(6\alpha^{2}\gamma(2\gamma-1))^{m}}{2(\alpha t+\beta)^{2m}}-\frac{3m\alpha^{2}\alpha_{1}(3m^{2}+3\gamma^{2}+2m\gamma-7m-4\gamma+4)}{(\alpha t+\beta)^{5-3m}}}}{+\frac{4\alpha^{2}\gamma^{2}\alpha_{2}(2m^{2}-4m^{2}\gamma-3m\gamma^{2}-6m+7m\gamma+6\gamma^{2}-6\gamma)}{(\alpha t+\beta)^{4-2m}} }}
{ \frac{m(1-m)(6\alpha^{2}\gamma(2\gamma-1))^{m}}{2(\alpha t+\beta)^{2m}}-\frac{4m(\gamma-1)\alpha^{3}\alpha_{2}\gamma^{2}(2\alpha\gamma^{2}+9m-9)}{(2\gamma-1)(\alpha t+\beta)^{5-3m}}-\frac{12\alpha^{3}\gamma^{3}\alpha_{2}(2m+m(\gamma-1)-2(\gamma-1))}{(\alpha t+\beta)^{4-2m}} }\right] .
\end{equation}

\begin{figure}[H]
\centering
\includegraphics[scale =0.5]{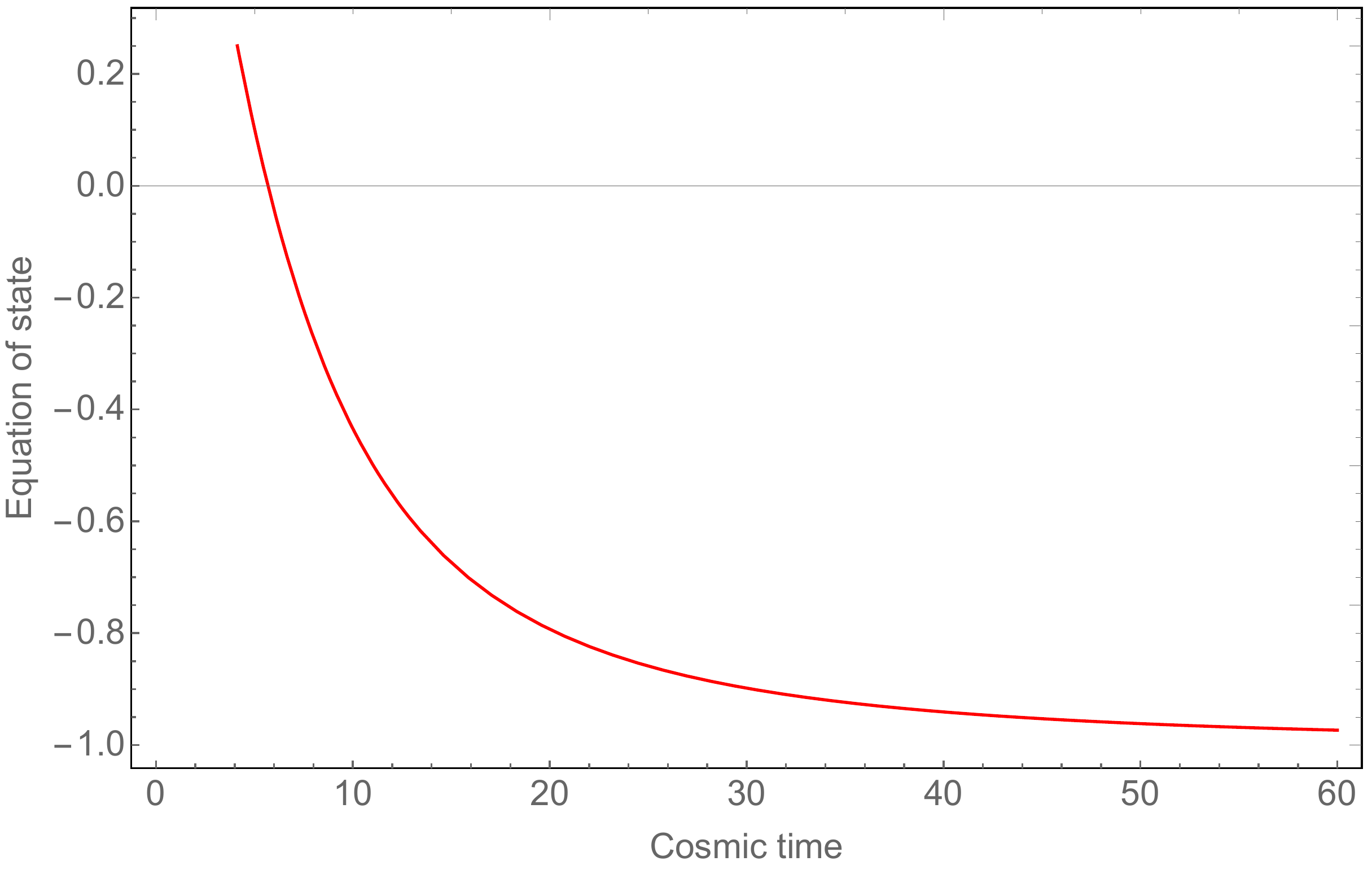}
\caption{Behavior of Equation of State parameter of the model versus cosmic time $t$ with $\alpha=0.9$, $\beta=4$, $\gamma=2.2$ $m=0.45$, $\alpha_{1}=0.073$, $\alpha_{1}=0.463$ .}\label{Figure-4}
\end{figure}

From the equation (\ref{e45}), it is observed that the EoS parameter of an isotropic model is a function of cosmic time $t$. The graphical behavior of EoS parameter verses cosmic time $t$ is shown in Figure {\ref{Figure-4}}. At the initial stage when the universe starts to accelerate for small interval of cosmic time $0.1\leqslant t \leqslant 5.5$,  the EoS parameter of the universe shows positive decreasing behavior which has a range $0.3944\leqslant \omega \leqslant 0.0153$.  It indicates that the model always start from radiation dominated era while for the some interval of cosmic time $5.5<t<65.1$, it has  $\omega> -1$  i.e. it shows quintessence region and for the whole interval of cosmic time $65.1<t\leqslant \infty$, it is negative and attain value -1. Hence the universe expand with radiation dominated era and at late times it is a $\Lambda$CDM model. The cosmological constant cold dark matter $\Lambda$CDM model is the simplest model of the universe that describes the present acceleration of universe and fits with the present day cosmological data \cite{37}, which is a situation in early universe where the $\Lambda$CDM field dominated universe may be playing an important role of the EoS parameter. Therefore, the model is always in acceleration.\\
\textbf{Speed of light in the model:}\\
For the stability of corresponding solutions of the model, we should check the model is physically acceptable. For this, first
required that the velocity of sound should be less than the velocity of light i.e. within the range $ 0<v^{2}=\frac{\partial p}{\partial \rho}$. In our model, we obtained the sound speed as
\begin{eqnarray*}
v^{2}=\frac{\partial p}{\partial \rho}
\end{eqnarray*}
\begin{equation}\label{e46}
\displaystyle v^{2}=\left[\dfrac{\splitdfrac{{\frac{m^{2}(m-1)(6\alpha^{2}\gamma(2\gamma-1))^{m}}{(\alpha t+\beta)^{2m+1}}+\frac{3m\alpha^{2}\alpha_{1}(3m-5)(3m^{2}+3\gamma^{2}+2m\gamma-7m-4\gamma+4)}{(\alpha t+\beta)^{6-3m}}}}{-\frac{4\alpha^{2}\gamma^{2}\alpha_{2}(2m-4)(2m^{2}-4m^{2}\gamma-3m\gamma^{2}-6m+7m\gamma+6\gamma^{2}-6\gamma)}{(\alpha t+\beta)^{5-2m}}}}
{\frac{m^{2}(1-m)(6\alpha^{2}\gamma(2\gamma-1))^{m}}{(\alpha t+\beta)^{2m+1}}+\frac{4m(\gamma-1)(3m-5)\alpha^{3}\alpha_{2}\gamma^{2}(2\alpha\gamma^{2}+9m-9)}{(2\gamma-1)(\alpha t+\beta)^{6-3m}}+\frac{12\alpha^{3}\gamma^{3}\alpha_{2}(2m-4)(2m+m(\gamma-1)-2(\gamma-1))}{(\alpha t+\beta)^{5-2m}} }\right].
\end{equation}

\begin{figure}[H]
\centering
\includegraphics[scale =0.50]{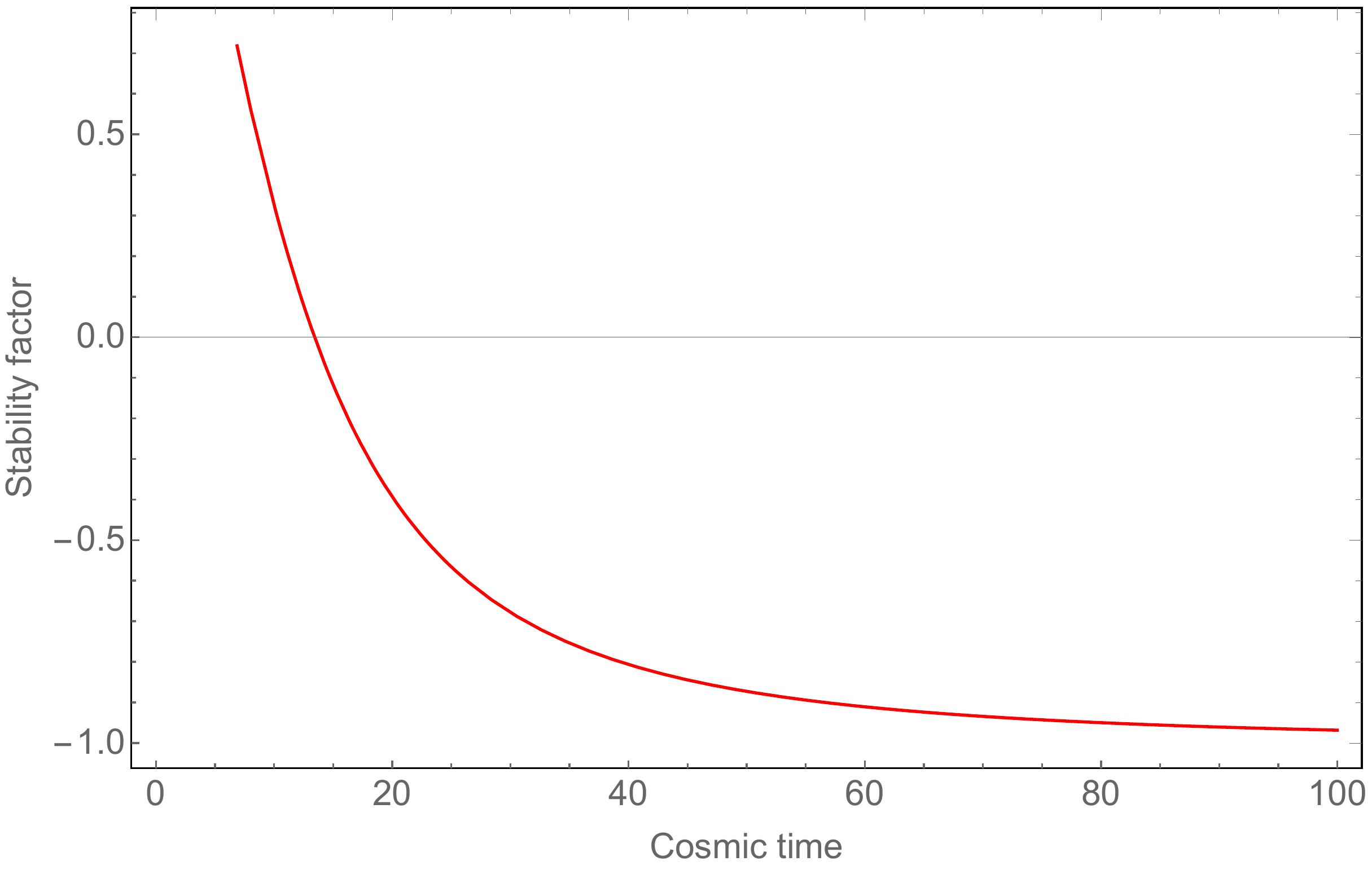}
\caption{Behavior of stability factor of the model versus cosmic time $t$ with $\alpha=0.9$, $\beta=4$, $\gamma=2.2$ $m=0.45$, $\alpha_{1}=0.073$, $\alpha_{2}=0.463$.}\label{Figure-iv}
\end{figure}

From Figure \ref{Figure-iv}, it is observed that stability factor behaves from positive to negative. Hence, initially $v^{2}>0$ in $0.1\leqslant t \leqslant 13.3$ which implies the model is stable and for $t> 13.3$, $v^{2}<0$, the derived model is unstable.

\subsection{Energy conditions}

In GR, the energy conditions (ECs) are a set of inequalities which describes the behavior of the compatibility of timelike, lightlike or spacelike curves. The ECs have significant theoretical applications, like the Hawking Penrose singularity conjecture, which is based on the strong energy condition (SEC) \cite{Hawking} whereas the dominant energy condition (DEC) is used to verify the positive mass theorem \cite{Schoen}. Further, the null energy condition (NEC) is a basic requirement to find the second law of black hole thermodynamics \cite{Carroll}. The generalized ECs have been studied in extended theories of gravity \cite{Capozziello} such as Brans-Dicke theory \cite{Atazadeh}, $f(R)$ gravity \cite{Santos}, $f(G)$ gravity \cite{Garcia}, $f(G,T)$ gravity \cite{Sharif} and $f(R,T)$ gravity \cite{Moraes,Yadav,Moraes2019}. 

The four different types of well known ECs are \cite{Visser}

Null energy condition: $\rho +p \geq 0$,

Weak energy condition: $\rho +p \geq 0$, $\rho \geq 0$,

Strong energy condition: $\rho +3p \geq0$, 

Dominant energy condition: $\rho \geq \mid p \mid$.

\begin{figure}[H]
\centering
\includegraphics[scale =0.50]{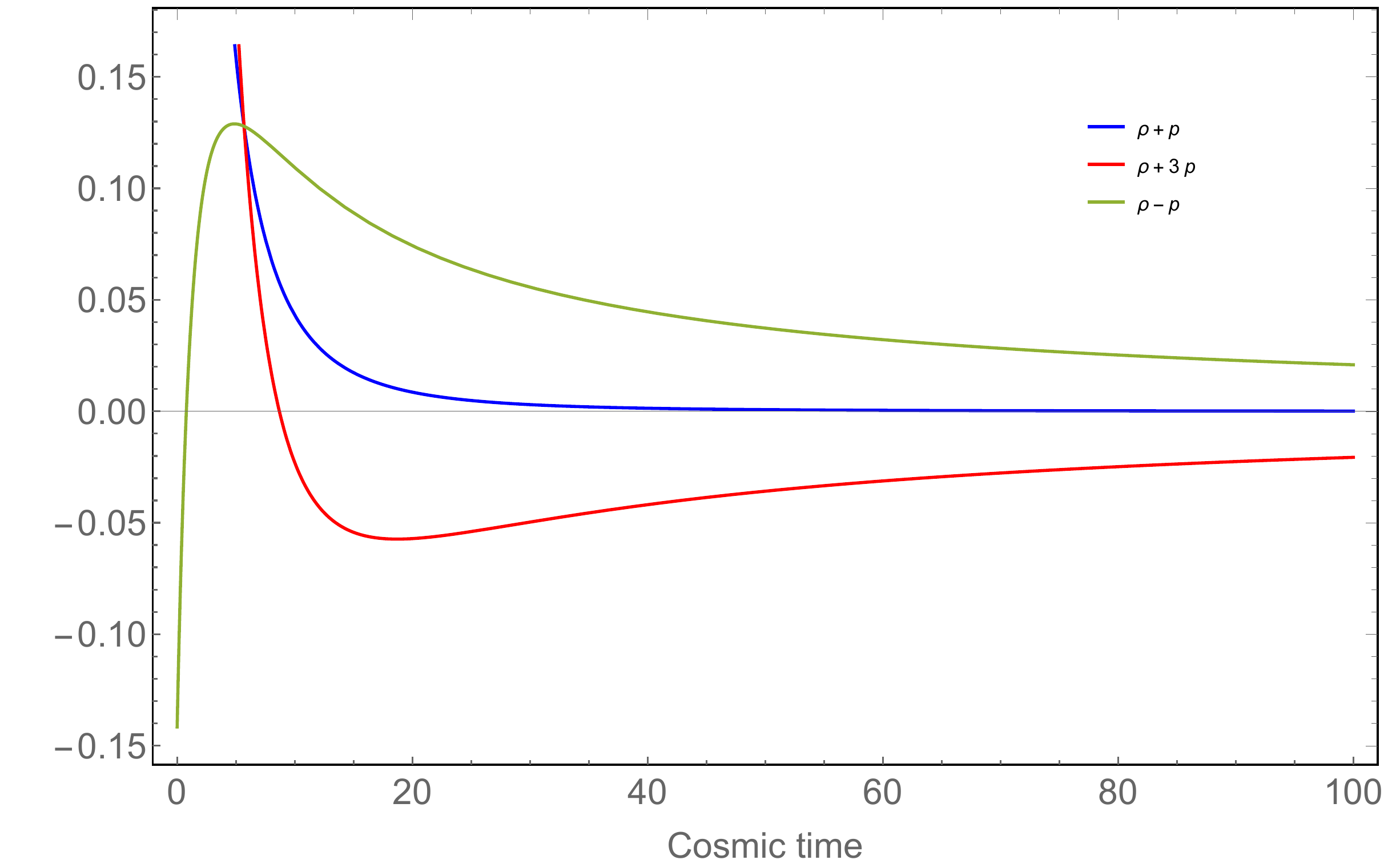}
\caption{Behavior of ECs of the model versus cosmic time $t$ with $\alpha=0.9$, $\beta=4$, $\gamma=2.2$ $m=0.45$, $\alpha_{1}=0.073$, $\alpha_{2}=0.463$.}\label{Figure-ec}
\end{figure}

We used the above relations for discussing the ECs in $f(R,G)$ gravity. The behavior of ECs in our model is presented in Figure \ref{Figure-ec}.

\subsection{Thermodynamical parameters of the model}

Temperature of this model is obtained as
\begin{equation}\label{e47}
\displaystyle \tau =\left[\splitdfrac{\frac{m(1-m)(6\alpha^{2}\gamma(2\gamma-1))^{m}}{2(\alpha t+\beta)^{2m}}-\frac{4m(\gamma-1)\alpha^{3}\alpha_{2}\gamma^{2}(2\alpha\gamma^{2}+9m-9)}{(2\gamma-1)(\alpha t+\beta)^{5-3m}}}
{-\frac{12\alpha^{3}\gamma^{3}\alpha_{2}(2m+m(\gamma-1)-2(\gamma-1))}{(\alpha t+\beta)^{4-2m}}}\right]^{\delta(t)}
\end{equation}
where
\begin{multline}
\delta(t)= \\ \dfrac{\left[ \dfrac{\left(\frac{m(m-1)[6\alpha^{2}\gamma(2\gamma-1)]^{m}}{2(\alpha t+\beta)^{2m}}-\frac{3m\alpha^{2}\alpha_{1}(3m^{2}+3\gamma^{2}+2m\gamma-7m-4\gamma+4)}{(\alpha t+\beta)^{5-3m}}+\frac{4\alpha^{2}\gamma^{2}\alpha_{2}(2m^{2}-4m^{2}\gamma-3m\gamma^{2}-6m+7m\gamma+6\gamma^{2}-6\gamma)}{(\alpha t+\beta)^{4-2m}}\right) }{\left( \frac{m(1-m)[6\alpha^{2}\gamma(2\gamma-1)]^{m}}{2(\alpha t+\beta)^{2m}}-\frac{4m(\gamma-1)\alpha^{3}\alpha_{2}\gamma^{2}(2\alpha\gamma^{2}+9m-9)}{(2\gamma-1)(\alpha t+\beta)^{5-3m}}-\frac{12\alpha^{3}\gamma^{3}\alpha_{2}[2m+m(\gamma-1)-2(\gamma-1)]}{(\alpha t+\beta)^{4-2m}}\right) }\right]} {\left[ 1+\dfrac{\left(\frac{m(m-1)[6\alpha^{2}\gamma(2\gamma-1)]^{m}}{2(\alpha t+\beta)^{2m}}-\frac{3m\alpha^{2}\alpha_{1}(3m^{2}+3\gamma^{2}+2m\gamma-7m-4\gamma+4)}{(\alpha t+\beta)^{5-3m}}+\frac{4\alpha^{2}\gamma^{2}\alpha_{2}(2m^{2}-4m^{2}\gamma-3m\gamma^{2}-6m+7m\gamma+6\gamma^{2}-6\gamma)}{(\alpha t+\beta)^{4-2m}}\right) }{\left( \frac{m(1-m)[6\alpha^{2}\gamma(2\gamma-1)]^{m}}{2(\alpha t+\beta)^{2m}}-\frac{4m(\gamma-1)\alpha^{3}\alpha_{2}\gamma^{2}(2\alpha\gamma^{2}+9m-9)}{(2\gamma-1)(\alpha t+\beta)^{5-3m}}-\frac{12\alpha^{3}\gamma^{3}\alpha_{2}[2m+m(\gamma-1)-2(\gamma-1)]}{(\alpha t+\beta)^{4-2m}}\right) }\right]} . 
\end{multline}

\begin{figure}[H]
\centering
\includegraphics[scale =0.50]{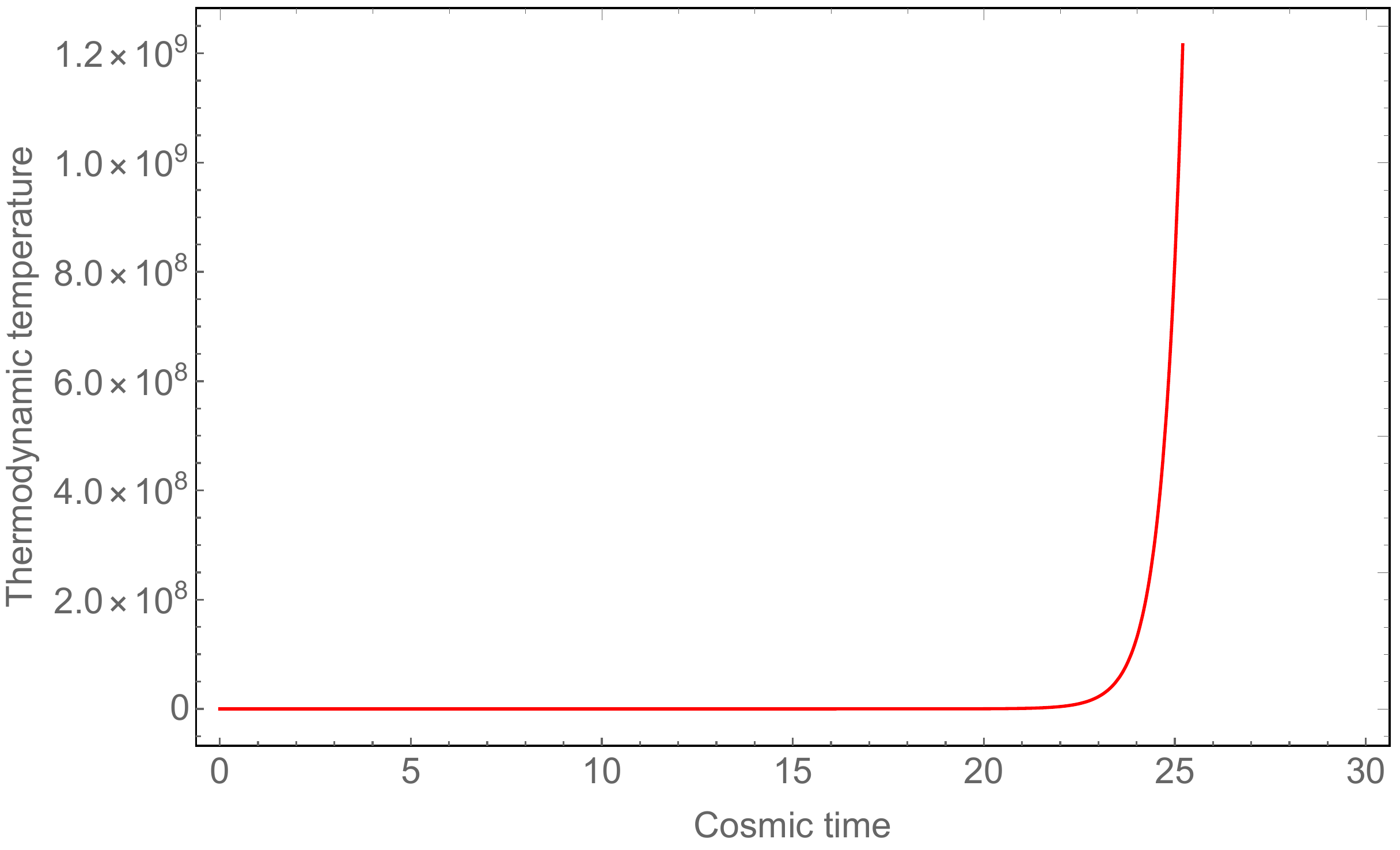}
\caption{Behavior of thermodynamic temperature of the model versus cosmic time $t$ with $\alpha=0.9$, $\beta=4$, $\gamma=2.2$ $m=0.45$, $\alpha_{1}=0.073$, $\alpha_{1}=0.463$ .}\label{Figure-v}
\end{figure}

From the equation (\ref{e47}), the thermodynamic temperature of the model is increasing with respect to expansion and at large expansion, it is infinite hence the model is not validating the second law of thermodynamic. The behavior of thermodynamic temperature with respect to expansion is clearly shown in Figure \ref{Figure-v} which resembles with the work in \cite{38,39}. 

The universe is filled with cosmic microwave background (CMB). Cosmic microwave background as
observed today consists of photons with an excellent black-body spectrum of temperature. The spectrum has been precisely measured by various instruments and does not show any deviation from the Planck spectrum \cite{28}.

Entropy density of the model is obtained as 
\begin{multline}\label{e48}
\displaystyle s'= \left[1+\dfrac{\splitdfrac{{\frac{m(m-1)(6\alpha^{2}\gamma(2\gamma-1))^{m}}{2(\alpha t+\beta)^{2m}}-\frac{3m\alpha^{2}\alpha_{1}(3m^{2}+3\gamma^{2}+2m\gamma-7m-4\gamma+4)}{(\alpha t+\beta)^{5-3m}}}}{+\frac{4\alpha^{2}\gamma^{2}\alpha_{2}(2m^{2}-4m^{2}\gamma-3m\gamma^{2}-6m+7m\gamma+6\gamma^{2}-6\gamma)}{(\alpha t+\beta)^{4-2m}}}}{ \frac{m(1-m)[6\alpha^{2}\gamma(2\gamma-1)]^{m}}{2(\alpha t+\beta)^{2m}}-\frac{4m(\gamma-1)\alpha^{3}\alpha_{2}\gamma^{2}(2\alpha\gamma^{2}+9m-9)}{(2\gamma-1)(\alpha t+\beta)^{5-3m}}-\frac{12\alpha^{3}\gamma^{3}\alpha_{2}[2m+m(\gamma-1)-2(\gamma-1)]}{(\alpha t+\beta)^{4-2m}} }\right]\times\\
\left[\splitdfrac{\frac{m(1-m)[6\alpha^{2}\gamma(2\gamma-1)]^{m}}{2(\alpha t+\beta)^{2m}}-\frac{4m(\gamma-1)\alpha^{3}\alpha_{2}\gamma^{2}(2\alpha\gamma^{2}+9m-9)}{(2\gamma-1)(\alpha t+\beta)^{5-3m}}}{-\frac{12\alpha^{3}\gamma^{3}\alpha_{2}[2m+m(\gamma-1)-2(\gamma-1)]}{(\alpha t+\beta)^{4-2m}}}\right] ^{\eta(t)}.
\end{multline}
where
\begin{equation}
\displaystyle \eta(t)= \dfrac{1} {1+\dfrac{\splitdfrac{{\frac{m(m-1)[6\alpha^{2}\gamma(2\gamma-1)]^{m}}{2(\alpha t+\beta)^{2m}}-\frac{3m\alpha^{2}\alpha_{1}(3m^{2}+3\gamma^{2}+2m\gamma-7m-4\gamma+4)}{(\alpha t+\beta)^{5-3m}}}}{+\frac{4\alpha^{2}\gamma^{2}\alpha_{2}(2m^{2}-4m^{2}\gamma-3m\gamma^{2}-6m+7m\gamma+6\gamma^{2}-6\gamma)}{(\alpha t+\beta)^{4-2m}} }} {\left( \frac{m(1-m)[6\alpha^{2}\gamma(2\gamma-1)]^{m}}{2(\alpha t+\beta)^{2m}}-\frac{4m(\gamma-1)\alpha^{3}\alpha_{2}\gamma^{2}(2\alpha\gamma^{2}+9m-9)}{(2\gamma-1)(\alpha t+\beta)^{5-3m}}-\frac{12\alpha^{3}\gamma^{3}\alpha_{2}[2m+m(\gamma-1)-2(\gamma-1)]}{(\alpha t+\beta)^{4-2m}}\right) }}
\end{equation}

\begin{figure}[H]
\centering
\includegraphics[scale =0.5]{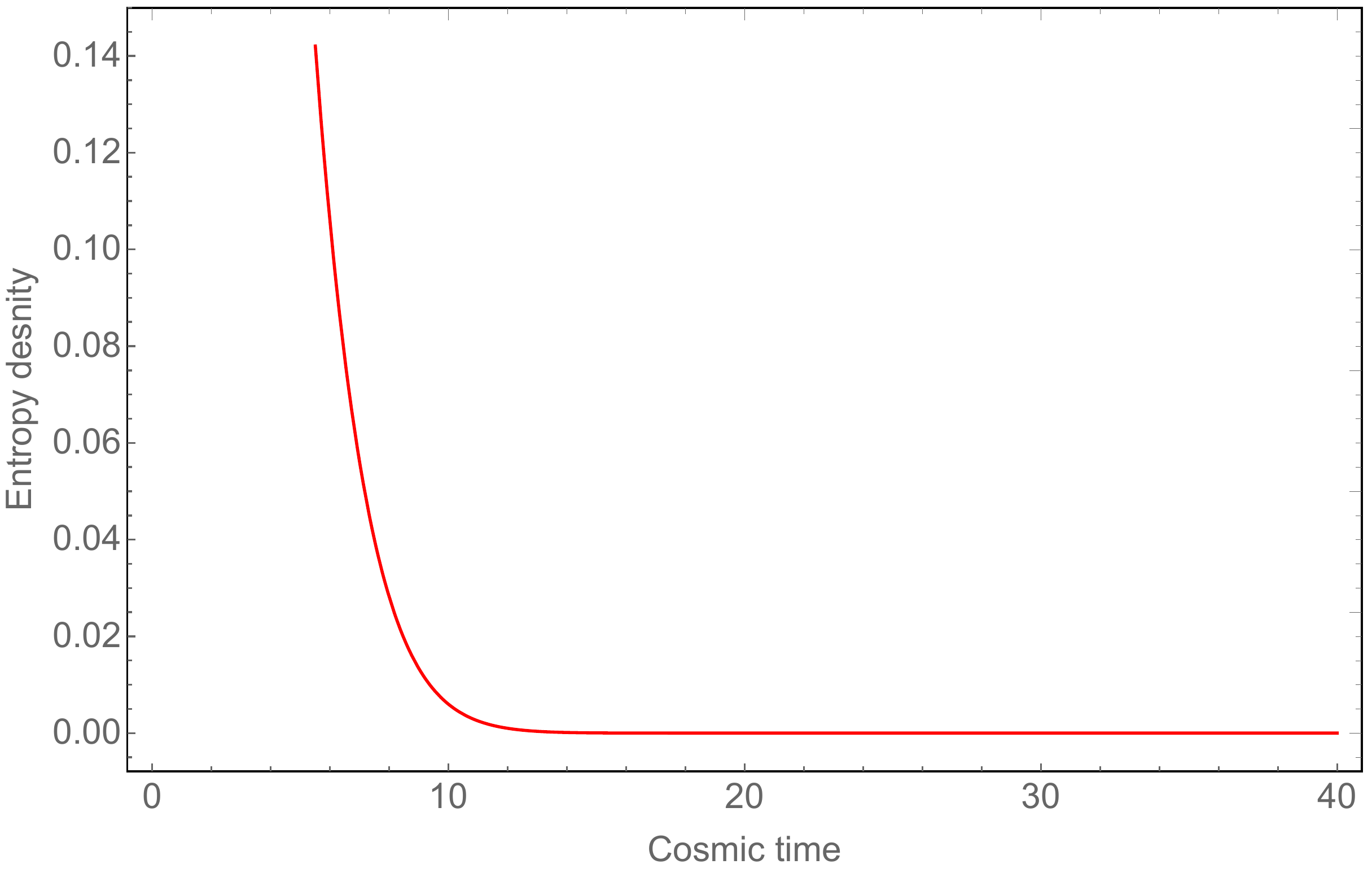}
\caption{Behavior of entropy density of the model versus cosmic time $t$ with $\alpha=0.9$, $\beta=4$, $\gamma=2.2$ $m=0.45$, $\alpha_{1}=0.073$, $\alpha_{2}=0.463$.}\label{Figure-vi}
\end{figure}

Equation (\ref{e48}) provide the entropy density of the model. As the entropy density of the model is directly related to the energy density with power term of equation of state parameter of the model, hence it shows slightly different behavior as that of energy density with the expansion. At infinite expansion, the thermodynamic energy density of the model is zero as seen in Figure {\ref{Figure-vi}}.

\section{Conclusion}\label{VI}

The present work was devoted to study the expansion with quintessence dark energy fluid with a solution based on constant deceleration parameter from the thermodynamic point of view for an isotropic and homogeneous cosmology.
In our investigations, the derived model shows initially different behavior as $t\rightarrow 0$.  The spatial volume of the model starts with constant value whereas it attains with big-bang as $t\rightarrow \frac{-\beta}{\alpha}$ and with the increase of cosmic time $t$ it always expands and increases, thus inflation is possible in flat isotropic FRW universe. The expansion scalar and the Hubble parameter both having an inverse relation with cosmic time. At big-bang both parameters have no identities and become constant throughout the evolution of the model as $t\rightarrow\infty$ which shows that the universe is expanding with the increase of cosmic time but the rate of expansion decreases to zero. Hence, the model starts evolving with zero volume as $t\rightarrow\infty$ with an infinite rate of expansion.

The energy density distribution of the model is positive decreasing function of cosmic time $t$. As $t\rightarrow 0$ the energy density of the model is constant whereas with the expansion as $t>0$ it is infinite and at the infinite time ($t\rightarrow\infty$) it approaches to zero i.e. $\rho\rightarrow 0$, thus at infinite expansion the model is asymptotically empty.

One can observe from Figure \ref{Figure-ec} that NEC, WEC, DEC are validated in our model whereas SEC is violated. The violation of SEC in our model indicates accelerated expansion of the presented model as discussed in Ref. \cite{Barcelo}.
 
At the initial stage when the universe starts to accelerate for small interval of time $0.1\leqslant t \leqslant 5.5$ the EoS parameter of the universe shows positive decreasing behavior which is a situation of radiation dominated era while for some interval of cosmic time it shows quintessence region. For the entire interval of cosmic time it is $\Lambda $CDM model, which is a situation in early universe where the $\Lambda $CDM model dominated universe may be playing an important role of the EoS parameter. The thermodynamic temperature of the model is increasing with respect to expansion and at large expansion, it is infinite. The behavior of temperature is in agreement with the standard thermodynamics for a homogeneous fluid.  The entropy density of the model is directly related with the energy density of the model. In the proposed model the entropy shows similar behavior as that of energy density with the expansion. Therefore, both the quantities entropy and temperature are positive in our model. At infinite expansion, the thermodynamic energy density of the model is zero. Hence, the model is not bearing out the second law of thermodynamics. Also the model is stable initially while for complete expansion it is unstable. Hence, our derived results resembles with the recent observational studies which includes \cite{3}-\cite{9}.

\section{Acknowledgements}
PKS acknowledges CSIR, New Delhi, India for financial support to carry out the Research project [No.03(1454)/19/EMR-II Dt.02/08/2019]. We are very indebted to the  editor and the anonymous referee for illuminating suggestions that have significantly
improved our paper in terms of  research  quality as well as presentation.

\end{document}